# A Comparative Evaluation of Predominant Deep Learning Quantified Stock Trading Stategies

Haohan Zhang
*Independent Researcher*
Shenzhen, China
hanarcher@berkeley.edu

*Abstract*—Quantified Stock Trading refers to the technique of delegating the buying and selling of stock shares to machines running a programmed algorithm. Since the problem at hand is essentially that of making decisions based on the processing of vast amounts of data, various deep learning models naturally emerge as ideal candidates. This study first reconstructs three deep learning powered trading models and their associated strategies that are representative of distinct approaches to the problem and established upon different aspects of the many theories evolved around deep learning. It then seeks to compare the performance of these strategies from the perspectives of fully informed vs. projection models, returns, risk vs. reward as well as similarity to the benchmark's return sequence's patterns through trading simulations ran on three scenarios when the benchmarks are kept at historical low points for extended periods of time. The results show that in extremely adverse market climates, investment portfolios managed by deep learning powered algorithms are able to avert accumulated losses by generating return sequences that shift the constantly negative CSI 300 benchmark return upward. Among the three, the LSTM model's strategy yields the best performance when the benchmark sustains continued loss.

## I. INTRODUCTION

Compared with the traditional approach of shareholders entrusting their investment portfolios to professional stock brokers, the best of whose performance might still be checked by the limitations of human faculties, the quantified trading machines can process vast amounts of information almost instantaneously while managing hundreds of kinds of stock shares simultaneously. In recent years online platforms such as JoinQuant [1] and BigQuant [2] have been introduced where users can construct their own machine learning powered algorithms, simulate trading scenarios based on historical market data, and ultimately employ their algorithms in actual real time trading. While the schools of thought on this matter are multifarious, after a period of exchange, debates and incorporations, there are some methods where consensus within the quantified trading community gradually converge on. The first step of this study is to analyze and reconstruct two models that are concretizations of some of these schools of thought, then apply these models in simulated trading using the JoinQuant platform. Subsequently, it introduces the architecture of another new model inspired by the previous ones. The performances of all three are compared with the aim of attempting at gaining perspectives on the most optimal way of harnessing the advantages of deep learning in the context of stock trading.

The stock market can be a massive, multi-dimensional and convoluted web of intricate information. This trait can have a somewhat deterring effect on researchers not necessarily conversant on the topics of financial engineering. In some cases, such problems can be alleviated by the introduction of factors. On a given trade day, the information accessible about any stock has many fields such as valuation, balance, and cash flow, etc., which in turn have their own sub-fields. Factors are results of established procedures, treatments and functions that convert select sub-fields of information into features that can be readily used as inputs into neural networks. These procedures are often validated by experts who can show that the fluctuation of these factors, while not necessarily prescient, is nonetheless closely related to a stock's future changes in certain aspects. One key factor around which considerable research has been generated is the relative strength indicator (RSI) invented by Wilder in 1978 [3]. Wilder uses an exponential moving average algorithm (EMA) that synthesizes the RSI factor as a gauge for the stock's strength and potentials. Expanding on Wilder's studies, Rodriguez Gonzalez [4] devised a scheme utilizing neural networks to yield a more accurate estimate of the RSI by choosing the most optimal window for the EMA calculations. Wilder and Rodriguez-Gonzalez's works mark a significant divergence from those of other researchers such as Liu et al. [5] who try to directly prophesize a stock's future prices based solely on a history of its past prices. Instinctively, it would be most ideal to simply foretell the future prices and take actions in advance. However, there are quite a few obstacles standing in the way of achieving such a feat. Chief among those is the level of noise and random variables inherent in any collection of stock information. Liu in his work addresses this issue by first employing an autoencoder to filter the noise from the data before adopting a long short memory (LSTM) recurrent neural network (RNN) to make predictions. The resultant prediction curve on the test data is purported to have low training loss and high prediction accuracy. A close examination may help conclude that the model actually generated a curve that is the real data shifted to the right by one time step in that, on a given trade day, when the model is asked to predict a stock's price on the following day, it returns a value which is very similar to the price on the day before. As the price differences between consecutive days are generally moderated (i.e., if a stock's price exceeds or falls below the close price of the day before by more than a certain percent threshold, trade limits are applied), it is indeed possible for such a model to yield low loss on the test set despite having qualified application in practical quantified trading.

Acknowledging the difficulties with directly predicting future prices based solely on past prices, researchers such as Wilder and Rodriguez-Gonzalez shift their focus onto using synthesized indicators derived from feeding basic factors into neural networks as a reference for a stock's future potentials. Since the indicators for all stocks in consideration are derived with the same method, they can be easily and justifiably compared and cross-examined. Henceforth, an effective trading strategy can be built from first establishing a ranking of stocks based on their indicator values then buying in stocks with high potentials and selling those whose potentials are relatively low. Such schemes no longer place a heavy emphasis on the accuracy of the predicted future prices, so long as the difference between the stock's potentials are correctly and proportionally modelled by the difference in the synthesized indicators.

## II. THE LINEAR REGRESSION MODEL

The linear regression model is inspired by Rhodes-Kropf et al. [6] who in 2005 put forth such a way of modeling the market capitalization of a stock based on three factors (the logarithm of net asset $b$, net profit $NI$, and financial leverage $LEV$, $IND$ is a matrix denoting which industry the stock in question belongs to, not counted as a factor, $m$ represents the logarithm of capitalization):

$$m = \alpha_0 IND + \alpha_1 b + \alpha_2 ln(NI)^+ + \alpha_3 I_{(<0)} ln(NI)^+ + \alpha_4 LEV + \varepsilon \quad (1)$$

Reiterating statements made in the previous section, the key point for this observed relationship in the context of quantified trading strategy is not its ability to reconstruct the logarithm of the market capitalization. This equation generalizes a descriptive relationship, not a predictive one. As the items in the equation above are the characteristics of a stock collected at the same time instance, it would be erroneous to try to use information on the right-hand side collected on one time instance and try to predict the item on the left-hand side of another. What can instead be harnessed from this relationship is the skewness term $\varepsilon$ which describes the difference between the modeled $m$ value and the actual observed data.

$$m_{predicted} = a_0 IND + a_1 b + a_2 ln(NI)^+ + a_3 I_{(<0)} ln(NI)^+ + a_4 LEV \quad (2)$$

$$\varepsilon = m_{predicted} - m_{actual} \quad (3)$$

The theoretical justification for this model is that for a normatively behaving stock, on any given trade day, the predicted $m$ should not have a significant skewness from the actual one. A large negative $\varepsilon$ value therefore implicates that at the particular time instance the stock is being underrated, which can lead one to reasonably conclude that it has high potentials for a rise in price in the near future. The strategy based on the linear regression model first collects factors (items on the right-hand side of (2)) and the logarithm of the market capitalization value (the $m$ value on the left-hand side of (2)) then adopts linear regression to numerically find the most suitable set of weights ($\alpha_1$, $\alpha_2$, $\alpha_3$) which maintains the mean squared error loss at a minimum on the training set. How far back to reach into history to construct the training set is a hyper-parameter that can be adjusted. All trade, as well as data collection occur on last trade days of a given month, henceforth called "action days".

As a concrete example, if the training set time window is set as 3 months, then on June 30 2015 (the last trade day of that month) information about all eligible stocks are collected on three preceding action days (March 31, April 30 and May 29) and used as training data. On an action day within the training set window, the model attempts to fit a linear combination of the available factors derived on that day to the stocks' true logarithm of market capitalization values. After the weights of the linear regression model have been learned, they are stored and used to derive $m_{predicted}$ values for all eligible stocks on June 30. The skewness values for all eligible stocks can be then calculated based on (3) and ranked. Stocks marked with low potentials are sold whereas those with high potentials are kept or acquired. No action or trading is performed until July 31, 2015, when the days from which to collect training set data instead becomes April 30, May 29 and June 30 2015. The strategy repeats this procedure, thus at any time within the duration of operation only keeping the stocks with the highest potentials, until a threshold return is achieved or the user manually terminates the execution. Figure 1 gives a flowchart of this strategy in fuller details.

Such a class of training method is named "rolling window" in the literature of quantified trading. On each last day of a month, information about the earliest day in the "window" is discarded while information about a new day is added to the training set. On each new action day, the strategy uses a new model trained by a set of data, a third of which is different from the one used by the model from the previous action day. This approach, to a certain degree, contradicts the traditional practices which advocate for one model trained on as much data, covering as many cases as possible, making decisions for all test samples. Indeed, in most applications, the test set accuracy of a model trained on a very abundant training set far exceeds that of a model trained on a scarcer one. This divergence is explained by the unique traits of the stock market as compared to other cases where deep learning models are applied. In most other cases, the mechanisms, conditions and settings of the problem tend not to change along the passage of time (invariant along time). For example, in speech recognition, a sentence spoken on one day presents the same features and patterns as one spoken on the following day. Therefore, congregating many speeches spoken across many intercepts along the time axis into a huge training set should present no difficulty. Under the influence of noise and random events, one model that learned certain patterns of the stock market during one period of time might not necessarily perform well on another period. One can only reach so far back into history when training a model before the collected data starts to become irrelevant as the latent variables governing the conditions of the stock market are increasingly different the further one goes back along the time axis. The "rolling window" approach offers one way of alleviating the issue by striving for a balance between the abundance and relevance of the data.

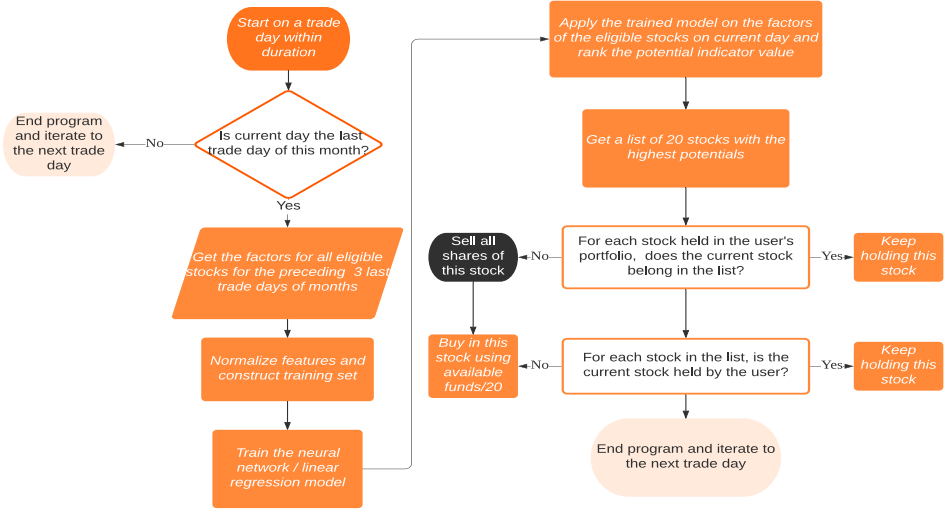

Fig.1. A flowchart for managing an investment portfolio based on the linear regression model

Additionally, when synthesizing the potential indicator, one needs not make the selection of factors to be exactly the same as those that appear in (1). Other factors may be substituted into the equation if proven propitious. For consistency, this study uses mostly the same collection of factors for all three models (the logarithm of market capitalization is a factor for the fully connected neural network and LSTM model, but it is the training label for the linear regression model, therefore the linear regression model has one less factor than the rest). 5 sample factors among the 47 about any single stock used in deep learning models discussed in this study are given in Table I (full list attached in appendix)

TABLE I. 5 SAMPLE FACTORS SHARED BY THREE MODELS

| Factors | Name/Methods of Derivation |
|---|---|
| 1 | EP: Net profit divided by market capitalization |
| 2 | The natural logarithm of the stock's price |
| 3 | EP cut: The difference between net profit and non-recurring gain loss divided by market capitalization |
| 4 | BP: Net assets divided by market capitalization |
| 5 | SP: Operating revenue divided by market capitalization |

### III. THE FULLY CONNECTED NEURAL NETWORK MODEL

On a particular action day, the linear regression model collects factor data from preceding action days to train the model and feed the factor data from the current day to make predictions about the that day's logarithm of market capitalization values. It then makes judgments about the stocks' potentials for future rises based on the difference between the predicted values and real values. Such a model is trained by past information, makes predictions about the current available information and makes decisions based on how much the real data differs from predictions. Technical market analysis [7] believes that all the external influences that play a role in the change of prices within the stock market are already reflected by the factors of the stocks. If we could follow technical analysis' school of thought, we may also believe that upon any intercept in the time axis when the strategy based on the linear regression model is about to make trading decisions, it knows all information sufficient to account for the stock market's characteristics and the external influences which a certain portion of the stocks are subject to should there exist such mathematical representations that can accurately quantify these characteristics and influences. We may call such models "fully informed".

Due to their relatively more complex architectures, other types of neural networks are sometimes tasked with constructing models that take a bolder step than the fully informed models. Such models train neural networks based on past information and make predictions about the stocks' certain factors in the near future based on a congregate of current and past information. Judgements about the stocks' potentials are thus formed based on these predictions. One proposed name for such models is "projection models". The second model this work investigates, the fully connected neural network model, belongs to the category of projection models. Figure 2 illustrates the difference between fully informed models and projection models.

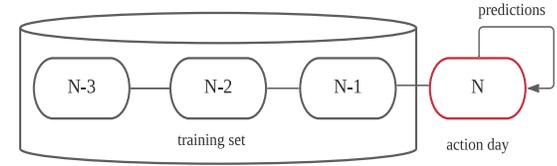

(a) Fully informed models

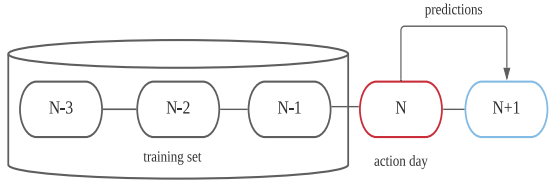

(b) Projection models

Fig. 2. Differences between (a) fully informed models and (b) projection models

Similar to the linear regression model, the fully connected neural network model also aims at synthesizing an indicator for potentials from input factors. Although two models share mostly the same factors as inputs, their methods with which to derive the training labels are actually different. On a particular action day, while the linear regression model fits the input factors against the same day's logarithm of market capitalization, the neural network model fits them against the excess return (the percent by which the stock in question's return exceeds that of the benchmark) on the last trade day of next month. For example, if the action day is again June 30 2015 then the linear regression model uses the market capitalization of the stock on June 30 2015 as the training label whereas the neural network model uses the stock's excess return on July 31 2015 with the factor data of the three previous action days as the training set. This model is named "projection model" because it is projecting the stocks' excess returns one month into the future. The architecture illustrated in Figure 3 is found to be ideal after a process of adjustments. The numbers of units on the hidden layers are respectively 32, 20 and 10. Rectified Linear Unit Activation (RELU) are applied on each layer except for the output layer since the label the neural network is fitting the output to is the excess return rate which is not necessarily a number between 0 and 1.

## IV. THE LSTM RECURRENT NEURAL NETWORK MODEL

The LSTM recurrent neural network model is another projection model. The two aforementioned models both collect data from three consecutive end of month action days to train a model for decision making. The factor information about each stock on each selected action day is treated as a parallel and independent data sample, assumed to have no correlation whatsoever with other data samples. This could be a reasonable assumption if the data samples are collected from different stocks on the same or different days. However, in the case of factors of the same stock collected from consecutive action days, such an approach would fail to capitalize on the sequential nature of the data which could potentially contain some latent features highly related to the stock's growth trends. Indeed, there have been several studies around the topic of projecting a record of a stock's prices along the axis of time as sequential data and using variations of a recurrent neural network to predict the stock's future prices. One such variation particularly favored by researchers in relevant fields is the Long Short Memory Unit (LSTM) which has a pronounced edge in preserving features from the earlier parts of a sequence for the reference of the units in the latter parts of the sequence. The schematic of a generic recurrent neural network which contains three time-steps is illustrated in Fig. 4. A recurrent neural network usually has one memory cell corresponding to each time-step within the sequence of input data. The rule of forward propagation between the memory cells are as follows:

$$a_t = g(W_{aa}a_{t-1} + W_{ax}x_t + b_a) \tag{4}$$

$$y_t = g(W_{ya}a_t + b_y) \tag{5}$$

$W_{aa}$, $W_{ya}$, and $W_{ax}$ are weight matrices while the two $b$s are biases. As the name of this type of neural network suggests, the same weights and biases are shared and used recurrently by all time-steps throughout the entire forward propagation. The circles within the memory cells represent the hidden units, the amount of which dictates the dimensions of the weight matrices.

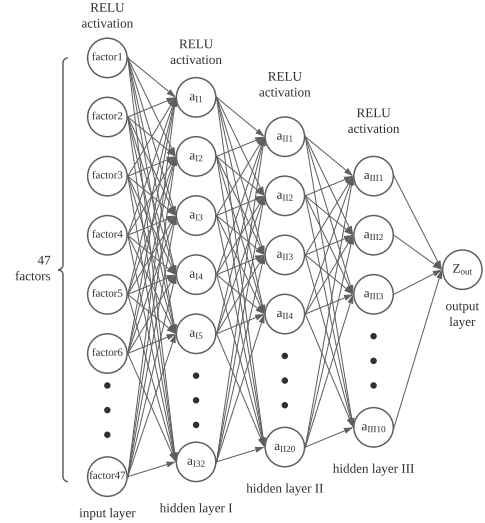

Fig. 3. The architecture for the fully connected neural network model

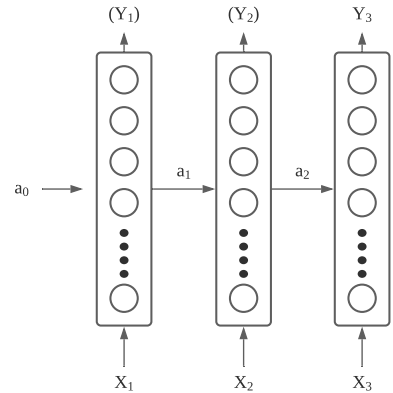

Fig. 4. A cerebral schematic for a generic recurrent neural network

The *g* functions are activation functions, their type to be specified depending on the application and other requirements. When designing the architecture of recurrent neural networks, one can choose to either keep all the output *y* values, thus generating a sequence of the same dimensions as the input or only keep the last *y* value. The sequence return is useful for language models or filtering the noise of a sequential data whereas the single *y* return is ideal for generating opinions, evaluations or categorizations based on the input data.

LSTM [8] adds modifications on top of the generic recurrent neural network model. The motivation comes from the observation that although the activated states of the hidden units are propagated through each layer, many important features about the inputs from the early time-steps are lost and not taken into account by the memory cells later on in the sequence. The LSTM model addresses this issue by adding a memory unit, $c_t$, and propagating it alongside $a_t$ with the hope that the memory unit would be able to retain these features from earlier time-steps. The algorithm for recurrently calculating the $c_t$ and $a_t$ values of a time-step based on the input and the $c_{t-1}$ and $a_{t-1}$ values from a time-step before is as follows:

$$\tilde{c}_t = tanh(W_{ca}a_{t-1} + W_{cx}x_t + b_c) \quad (6)$$

$$\Gamma_u = \sigma(W_{ua}a_{t-1} + W_{ux}x_t + b_u) \quad (7)$$

$$\Gamma_f = \sigma(W_{fa}a_{t-1} + W_{fx}x_t + b_f) \quad (8)$$

$$\Gamma_o = \sigma(W_{oa}a_{t-1} + W_{ox}x_t + b_o) \quad (9)$$

$$c_t = \Gamma_u * \tilde{c}_t + \Gamma_f * c_{t-1} \quad (10)$$

$$a_t = \Gamma_o * tanh(c_t) \quad (11)$$

Similar to the case of a generic RNN model, *W*s are weight matrices, *b*s are bias terms, all of which are learned through back propagations. $\sigma$ and *tanh* are sigmoid and hyperbolic tangent activation functions. The LSTM model introduces new gate components $\Gamma_u$, $\Gamma_f$, and $\Gamma_o$ which are intermediary values that play a determinant role in the final assignment of the $a_t$ and $c_t$ values. A cerebral schematic of the LSTM architecture applied in this study is illustrated as follows.

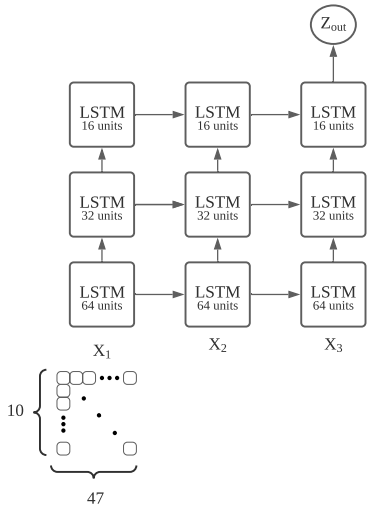

Fig. 5. The architecture of the LSTM model used in this study

On any action day, the factor data for three previous consecutive action days form the input X sequence which has three steps. As the batch size is set to 10 and the total number of factors is 47, the dimension of the input of each time-step is 10 by 47. The first two layers of the LSTM network return sequences while the last one returns a single value $Z_{out}$, which, similar to the procedure in the fully connected neural network model, is fitted to the excess return of the next action day. For example, if the action day in question is again Jun 30 2015. For each eligible stock, its factors on March 31, April 30, and May 29 2015 forms a sequential training sample, and its excess return on Jun 30 2015 would be the training label that $Z_{out}$ is fitting to. After training is completed, the model predicts the excess returns of eligible stocks based on the factor data from April 30, May 29, and Jun 30 2015, a time window that is the training set's time window shifted to the right by one action day. This set of predictions is the basis for making judgements about the stocks' potentials and the subsequent trade procedures which are similar to the ones illustrated by Figure 1. After all actions have been successfully performed, the strategy hibernates and waits until the next last trade day of the month, July 31 2015 to be activated when, similar to before, the data from April 30, May 29, Jun 30 2015 would be used to train the model and data from May 29, Jun 30, and July 31 2015 would be used to project the excess return on August 31.

After 10 epochs of training, the mean squared error loss is brought down to less than 0.01, Figure 6 shows the prediction made by the LSTM model of the eligible stocks' excess returns on Jun 30 2015 overlaid by that day's true values. One observation that can be made from Figure 6 is that the LSTM network has a moderating effect on the extreme and noisy values of the original y values.

## V. COMPARISONS UNDER SIMULATIONS

When gauging the performance of a certain trading strategy, more important than how it maximizes the return during robust market growths is its ability to staunch the loss or even achieve positive growths in very adverse market climates. Those strategies that are able to amplify the return when the whole market is doing well but forfeits those gains by yielding to drastic dives as soon as the benchmark fluctuates are obviously not desirable. It is during the adverse situations when the whole market sustains continual loss for long periods of time that the true reliability and steadfastness of a trading strategy can be tested.

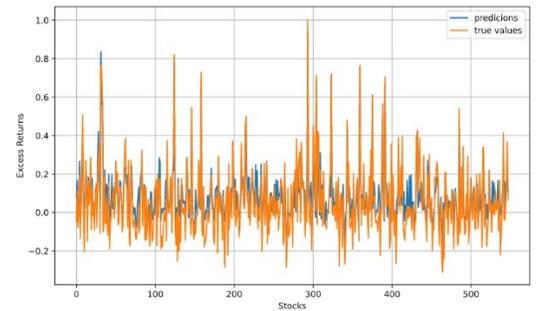

Fig. 6. The predictions of excess returns on Jun 30, 2015 and true values

This study runs trading strategies derived from the three aforementioned models on simulated scenarios based on three six-month periods when the benchmark was kept at historical low points. The performances of the portfolios managed by three strategies are evaluated from the aspects of aggressiveness, risk vs. reward as well as similarity in trends with the benchmark.

The risk vs. reward is measured by the Sharpe Ratio, which is developed by Sharpe [9] in 1966. It is defined as:

$$R_{sharpe} = (R_p - R_f)/\sigma_p \quad (12)$$

$R_p$ is the return of the strategy whereas $R_f$ is the risk-free return of the market (e.g., the interest rate of a conservative long-term bank deposit stipulated by the Federal Reserve), $\sigma_p$ is the standard deviation of excess returns, also called volitivity. A record of an investment portfolio's excess return compared with a benchmark on consecutive trade days constructs a time series, the standard deviation of which measures the degree of fluctuation of the portfolio's excess return. The amount by which the portfolio exceeds the risk-free return per unit of volitivity offers one way of quantitatively describing the risk vs. reward ratio. On the other hand, the similarity between the portfolio's daily return series and that of the benchmark is acquired by first taking the difference between the two series and taking the standard deviation of the difference. Intuitively, if one adds a constant offset to each element of the series to get a new series, the standard deviation of the difference between the new series and the original would be zero. Consequently, the higher the standard deviation, the more fluctuation there is between the difference of each corresponding element from the two series. In each of the following scenario, the sequence of daily returns of the CSI 300 [10] benchmark, a capitalization weighted index reflecting the prices of 300 major stocks on the Shanghai Stock Exchange and Shenzhen Stock Exchange, is also plotted in red to offer a reference as to the portfolios' strength as compared to other major investments in the market.

*A. Constant Negative Benchmark Returns with a Point of Mild Inflection*

The six-month period from June 1 2015 to January 1 2016 witnesses a continuous negative benchmark return which projects to an annual return rate of -13.73%. Simulated portfolios were created and managed by strategies based on each of the three models. A chart showing the characteristics of the daily return series of the portfolios is as follows:

TABLE II. CHARACTERISTICS OF RETURN SERIES FOR THREE PORTFOLIOS IN SCENARIO A

| Characteristics | Models | | |
|---|---|---|---|
| | Linear Regression | Fully Connected Neural Network | LSTM |
| Sharpe Ratio | 0.277 | -0.618 | -0.457 |
| Net Return | 8.58% | -8.26% | -7.9% |
| Benchmark Return | -4.69% | -4.69% | -4.69% |
| Similarity to Benchmark | 0.11 | 0.45 | 0.39 |

Their daily return sequences compared with that of the CSI 300 benchmark are plotted in Figure 7. Although the benchmark return was negative throughout this period, there was a point of mild inflection that occurred during October which enabled it to climb back from the lowest point of -37.5% on August 26 to a -4.69% at the end of the period. Table III offers an account of the benchmark return by month in order to illustrate this point of inflection:

TABLE III. CSI 300 BENCKMARK'S RETURN BY MONTH DURING SCENARIO A

| Jun | Jul | Aug | Sep | Oct | Nov | Dec |
|---|---|---|---|---|---|---|
| -0.08 | -0.15 | -0.12 | -0.05 | 0.10 | 0.01 | 0.05 |

It is clear that the monthly return inflected in October and the returns of the subsequent two months are both positive. During adverse market climates, one of the few ways for a strategy to still achieve positive return is to be extremely sensitive to the market's upward inflections and relatively impervious to its downward slides. In this scenario, the linear regression model has clear advantages over the other two in all aspects because of its markedly superior sensitivity to the rejuvenation in October. Namely, the mildest tick in the market caused a remarkable surge in the portfolio's return. A record of the Sharpe Ratio of the linear regression model's portfolio is given in Table IV:

TABLE IV. SHARPE RATIO OF LINEAR REGRESSION MODEL'S PORTFOLIO BY MONTH DURING SCENARIO A

| Jun | Jul | Aug | Sep | Oct | Nov | Dec |
|---|---|---|---|---|---|---|
| 4.00 | -1.55 | -0.95 | -0.99 | 118.72 | -0.46 | -0.49 |

One pattern that can be traced from this experiment is that in A-like scenarios, the more similar a strategy's portfolio is to the benchmark in trends of the return series, the poorer it performs. The portfolio corresponding to the fully connected neural network model has the highest similarity to benchmark value, but its return over the six-month period appears to be the lowest. For a portfolio that augmented on the upward inflection, its trend in the return series must have deviated from that of the benchmark. Conversely, for a portfolio that merely copied the inflection but did not capitalize on it, its trend must have been more similar.

*B. Negative and Fluctuating Benchmark Returns with a Continued Downward Slide*

The benchmark returns series during the period from June 1 2018 to January 1 2019 is highly fluctuating. It started with a -7.66% at the end of June, suffered a major dive in October and never truly recovered, resulting in a -20.82% at the end of the period. A full record of the benchmark's return by month is given in Table V:

TABLE V. CSI 300 BENCKMARK'S RETURN BY MONTH DURING SCENARIO B

| Jun | Jul | Aug | Sep | Oct | Nov | Dec |
|---|---|---|---|---|---|---|
| -0.08 | 0.00 | -0.05 | 0.03 | -0.08 | 0.01 | -0.05 |

Their daily return sequences compared with that of the CSI 300 benchmark are plotted in Figure 8. In comparison, the benchmark's return series is fluctuating more than that of Scenario A. There is no consistent trend to follow since surges are usually immediately followed by plunges, and vice versa. Table VI shows the characteristics of the daily return series of the portfolios.

TABLE VI. CHARACTERISTICS OF RETURN SERIES FOR THREE PORTFOLIOS IN SCENARIO B

| Characteristics | Models | | |
| --- | --- | --- | --- |
| | Linear Regression | Fully Connected Neural Network | LSTM |
| Sharpe Ratio | -1.35 | -1.98 | -1.13 |
| Net Return | -6.75% | -12.16% | -5.7% |
| Benchmark Return | -20.82% | -20.82% | -20.82% |
| Similarity to Benchmark | 0.46 | 0.50 | 0.59 |

In this scenario, the portfolio managed by the LSTM model shows advantages in all aspects.

*C. The Financial Crisis of 2008*

Another historically renown scenario that is similar to Scenario B is the financial crisis in 2008 which had wide and deep repercussions throughout the world's economy. From February to August of that year, the CSI 300 benchmark return dropped by an astounding 30%. In scenario C, the strategies based on the three models are run during this six-month period to test their resilience and robustness against continual heavy benchmark loss. The characteristics of the daily return series of the portfolios is shown by Table VII. Figure 9 also plots their daily return sequences together as compared with that of the benchmark.

Scenario C is, in essence, a more extreme version of Scenario B. The fact that the LSTM model continues to out-perform the other two models is consistent with the observations made from B.

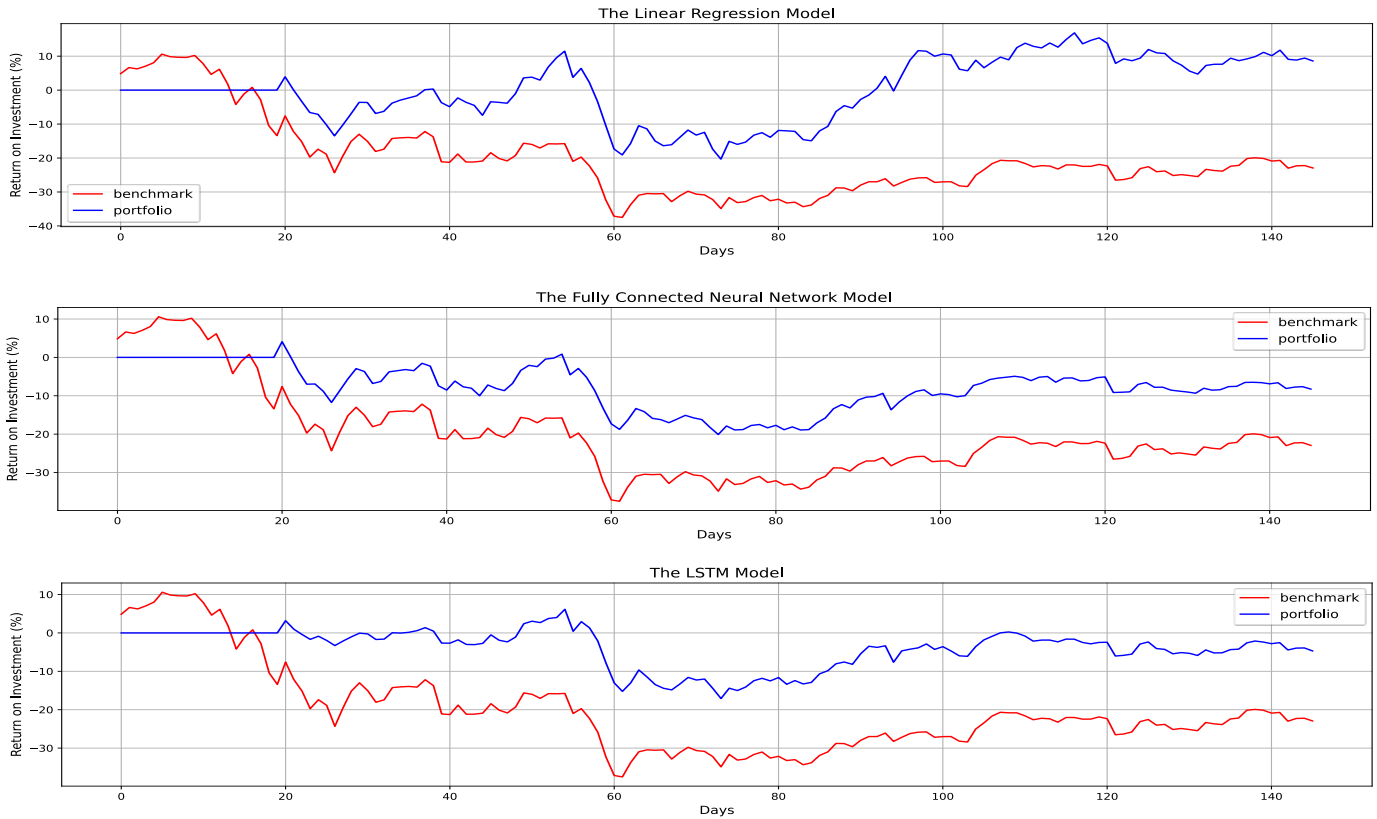

Fig. 7. The daily return sequences of portfolios managed by the three models during scenario A

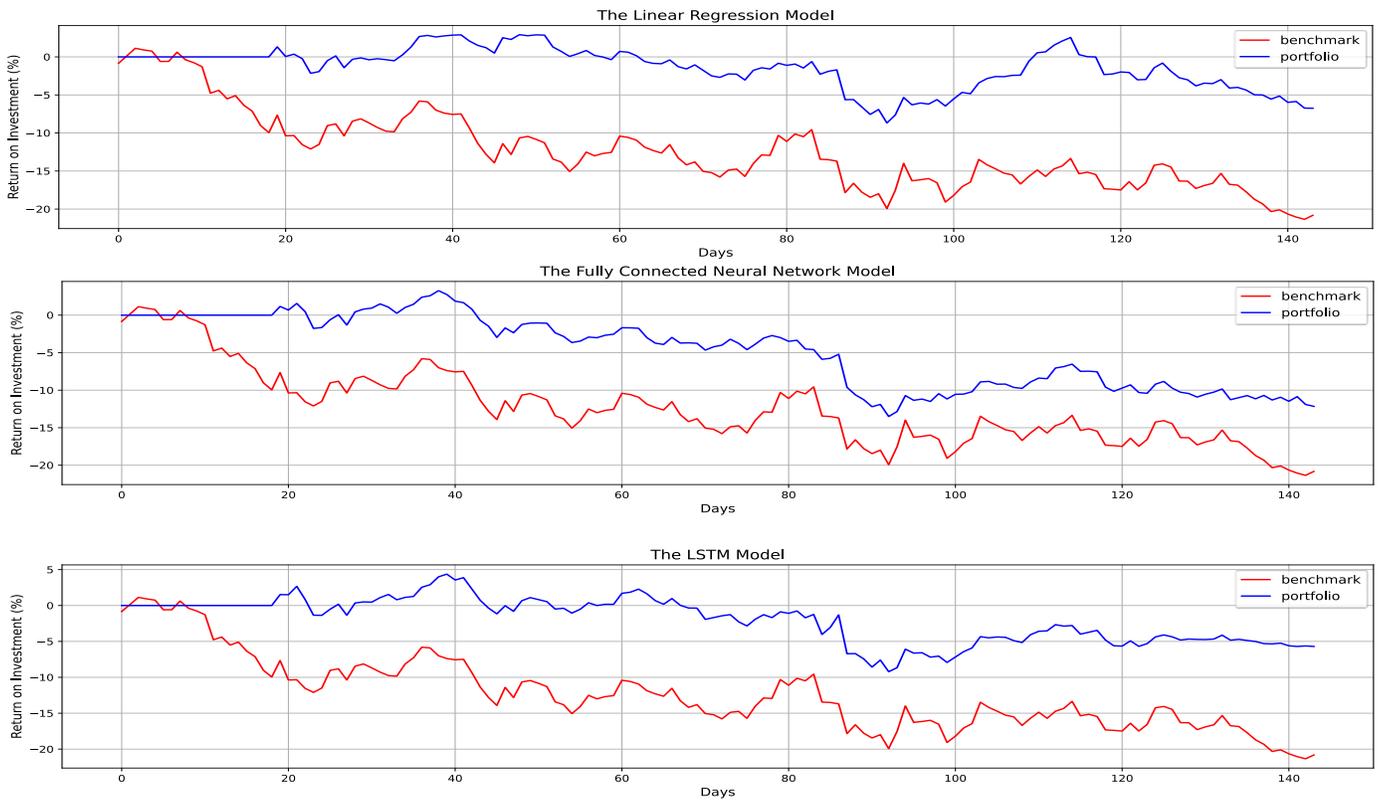

Fig. 8. The daily return sequences of portfolios managed by the three models during scenario B

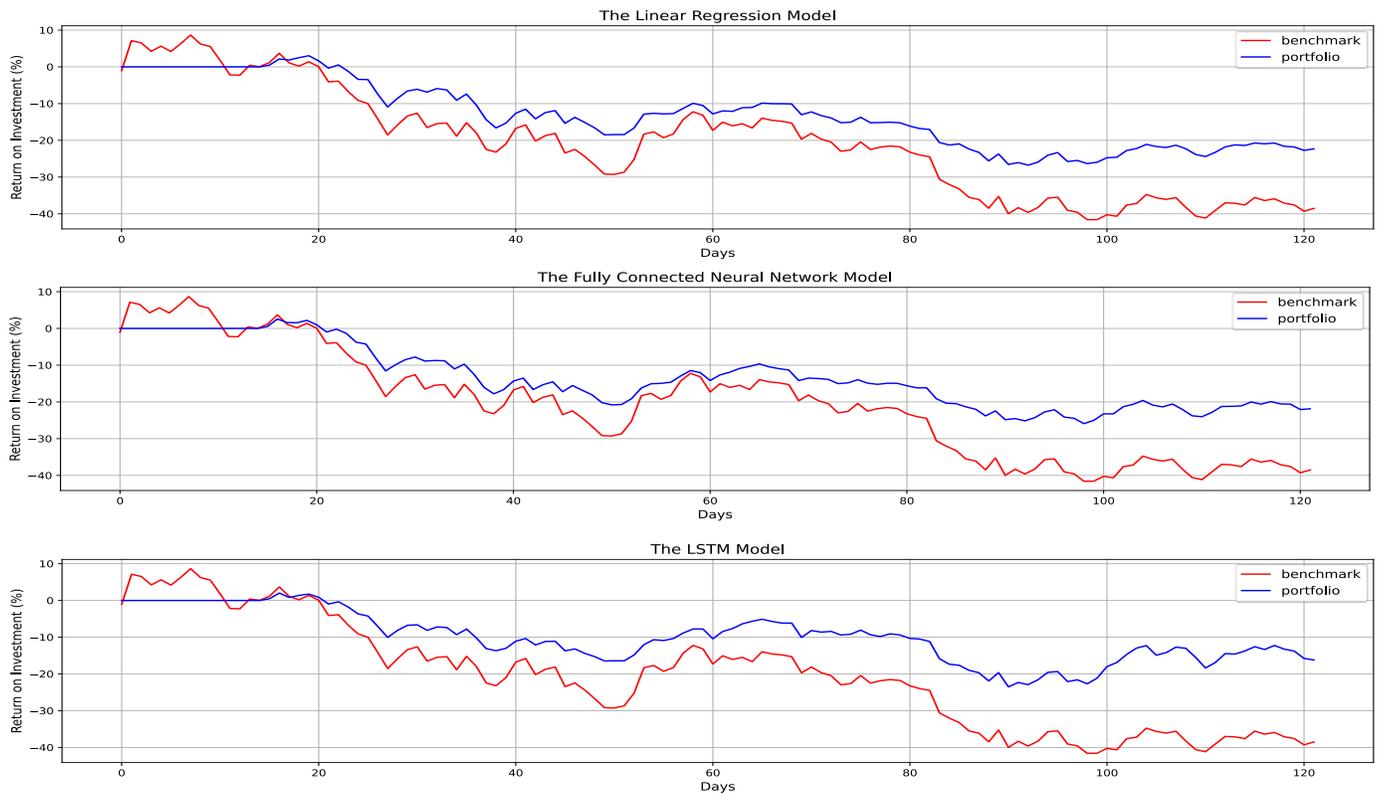

Fig. 9. The daily return sequences of portfolios managed by the three models during scenario C

TABLE VII. CHARACTERISTICS OF RETURN SERIES FOR THREE PORTFOLIOS IN SCENARIO C

| Characteristics | Models | | |
|---|---|---|---|
| | Linear Regression | Fully Connected Neural Network | LSTM |
| Sharpe Ratio | -1.69 | -1.77 | -1.27 |
| Net Return | -22.32% | -21.87% | -16.19% |
| Benchmark Return | -40.41% | -40.41% | -40.41% |
| Similarity to Benchmark | 0.22 | 0.18 | 0.15 |

Overall, the portfolio managed by the fully connected neural network model is ranked the lowest in terms of performance. Not only does it yield low returns, it also has relatively high risk given the same amount of reward. The linear regression model performs better when there is an inflection point in the market to be taken advantage of, but is less stable when the market fluctuates and sustains continued loss as in scenarios B and C. Pitting strategies based on projection models against those based on fully informed models is also a test of the projection model strategy's stability when put under unforeseen and unexpected influences. Although admittedly between the action day and the last trade day of the next month events disrupting the world's economy could affect the quality of the judgements, the LSTM still has a better grasp of the patterns in the temporal variation of the top-ranking stocks' daily returns than the linear regression model owing to recurrent neural networks' inherent superiority in learning from sequential data. On the other hand, while the linear regression model's strategy was able to excel at amplifying the upward inflection in scenario A, its total disregard of the sequential nature of the data led it to sustain a heavier loss and higher risk vs. reward ratio than the LSTM model's strategy who learned signs from the patterns of the series input indicative of continued negative returns that are incomprehensible to the linear regression model due to its architecture.

## VI. CONCLUSION AND REMARKS

In adverse market climates, deep learning powered automated trading strategies have the ability to manage investment portfolios whose sequence of daily returns resemble closely that of the CSI 300 benchmark shifted upward. This is generally achieved by using deep learning models trained by factor data from the past to make judgements about the stocks' future profit potentials and managing a portfolio that only retains those stocks with the highest potentials. Thus, even in extreme scenarios when the return of the benchmark keeps at alarmingly low values for a very extended period of time, these models can still help avert accumulative loss or even achieve substantial gains. Although these judgements about the stocks' profit margins can in no way be guaranteed to be completely accurate, overall, the loss incurred from incorrect ones is outweighed by the gain contributed by the sound ones. The linear regression model can be more aggressive in time periods when there is an inflection point in the benchmark's returns, but loses its edge to the LSTM model in steadily declining and vibrantly fluctuating markets. On the other hand, the fully connected neural network's ability to use hidden layers to detect features and characteristics of objects in an image does not render the most exceptional trading strategies where analysis of sequential data is involved.

From the perspective of fully informed models vs. projection models, it may be safely concluded that fully informing a model is more important if the goal is to amplify an uptick in the general market. On the other hand, if the goal is to keep loss at a minimum in B and C like scenarios, then taking into account the temporal and sequential nature of the data is more important.

APPENDIX: A FULL TABLE OF THE 47 FACTORS USED BY THE 3 MODELS

| Factors | Name/Methods of Derivation |
|---|---|
| 1 | EP: Net profit divided by market capitalization |
| 2 | The natural logarithm of the stock's price |
| 3 | EP cut: The difference between net profit and non-recurring gain loss divided by market capitalization |
| 4 | BP: Net assets divided by market capitalization |
| 5 | SP: Operating revenue divided by market capitalization |
| 6 | NCFP: Cash flow rate divided by market capitalization |
| 7 | OCFP: Net operate cash flow rate divided by market capitalization |
| 8 | G/PE: Net profit growth rate divided by PE ratio |
| 9 | ROE: Net income divided by equity |
| 10 | ROA: Net income divided by average total assets |
| 11 | Gross profit margin |
| 12 | Profit margin |
| 13 | Asset turnover rate |
| 14 | Operation cashflow ratio: Net operate cash flow divided by operate income |
| 15 | Financial leverage: Total assets divided by net assets |
| 16 | Debt equity ratio: Long term debt divided by net assets |
| 17 | Cash ratio |
| 18 | Current ratio |
| 19 | Logarithm of market capitalization |
| 20 | Return over one month |
| 21 | Return over three months |
| 22 | Return over six months |
| 23 | Return over twelve months |
| 24 | The average of the result of the element-wise multiplication of daily returns and daily turnover ratio over a period of 1, 3, 6 and 12 months |
| 25 | |
| 26 | |
| 27 | |
| 28 | The average of the result of the element-wise multiplication of the sequence of daily returns, daily turnover ratio and exp(-xi/N/4) over a period of 1, 3, 6 and 12 months, where xi is the distance in days between the action day and the day in the sequence and N is the number of months |
| 29 | |
| 30 | |
| 31 | |
| 32 | The standard deviation of the stock's sequence of daily returns over a period of 1, 3, 6, and 12 months |
| 33 | |
| 34 | |
| 35 | |
| 36 | The stock's turnover ratio over the past 1, 3, 6 and 12 months minus one |
| 37 | |
| 38 | |
| 39 | The stock's turnover ratio over the past 1, 3, 6 and 12 months divided by the stock's turnover ratio over the past two years minus one |
| 40 | |
| 41 | |
| 42 | |
| 43 | |
| 44 | beta [11] |
| 45 | MACD [12] |
| 46 | DEA [13] |
| 47 | DIF [13] |

## Authors' background

| Your Name | Title* | Research Field | Personal website |
|---|---|---|---|
| Haohan Zhang | Professional | Artificial intelligence used in the field of automation and stock markets | |
| | | | |
| | | | |
| | | | |